\documentclass
{mn2e}
\usepackage{epsfig}
\usepackage{amsmath, amssymb,bm}

\def \be{\begin{equation}}
\def \ee{\end{equation}}

\def \msun{\rm M_{\odot}}

\begin{document}
%\title[]{ The Perils of Pdot}
%\author[Andrew King \& Jean--Pierre Lasota] 
%{\parbox{5in}{Andrew King$^{1, 2, 3}$ \& Jean--Pierre Lasota$^{4, 5}$
%%%%%%%%%%%%%%%%%%%%%%%%%%%%%%%%%%%%%%%%%%%%%%%%%%%%%%%%%%%%%%%%%%%%%%%%%%%%%%
%% Title Details and Page Header                                            %%
%%%%%%%%%%%%%%%%%%%%%%%%%%%%%%%%%%%%%%%%%%%%%%%%%%%%%%%%%%%%%%%%%%%%%%%%%%%%%%
\title[] {The Rapidly--Changing Period of the QPE Source 1ES~1927+654
}

\author[Andrew King] 
{\parbox{5in}{Andrew King$^{1, 2, 3}$ 
}
\vspace{0.1in} \\ $^1$ School of Physics \& Astronomy, University
of Leicester, Leicester LE1 7RH UK\\ 
$^2$ Astronomical Institute Anton Pannekoek, University of Amsterdam, Science Park 904, NL--1098 XH Amsterdam, The Netherlands \\
$^{3}$ Leiden Observatory, Leiden University, Niels Bohrweg 2, NL--2333 CA Leiden, The Netherlands}
\maketitle
\begin{abstract}
Several low--mass galaxy nuclei are observed to produce quasiperiodic eruptions (QPEs). Recently one of these systems, 1ES~1927+654, changed its quasiperiod drastically, from $\sim 18$ minutes to $\sim 7.1$ minutes, over a span of just two years. I suggest that this is an effect of von Zeipel -- Lidov -- Kozai (ZLK) cycles, where a more distant star orbits the QPE `binary' in which a white dwarf orbits a moderately massive central black hole.  
%%%%%%%%%%%%%%%%%%%%%%%%%%%%%%%%%
I show that in 1ES~1927+654 the white dwarf's orbital
plane oscillates with angular amplitude $\simeq 71^{\circ}$ each side of the orbital plane of the distant star. This causes correlated changes of the orbital eccentricity and quasiperiod, and of the accretion luminosity driven by gravitational radiation losses. 
The GR luminosity has the characteristic property that it is inversely proportional to the instantaneous binary quasiperiod in all cases. The QPE system is probably just one of the effects produced by a complex infall event involving several stars. The whole system is likely to evolve rapidly, and will repay further monitoring.
%%%%%%%%%%%%%%%%%%%%%%%%
\end{abstract}

\begin{keywords}
{galaxies: active: supermassive black holes: black hole physics: X--rays: 
galaxies}
\end{keywords}

\footnotetext[1]{E-mail: ark@astro.le.ac.uk}
\section{Introduction}
\label{intro}

%Accreting systems frequently produce quasiperiodic luminosity variations. 
Observations show that a significant number of low–mass galaxy nuclei produce quasiperiodic eruptions (QPEs) in ultrasoft X–rays. These systems emit most of their  accretion luminosity in short but very bright bursts recurring at slightly varying intervals. These are of order a few hours in most cases, with two possible candidates at much longer quasiperiods.
 
Recently, Masterson et al. (2025) found that in 1ES~1927+654, which has a black hole mass of $\simeq 1.38\times 10^6\msun$, the quasiperiod $P$ shortened from an initial value $\sim 18$~min down to $\sim 7.1$~min over two years, with decelerating period evolution (i.e. $\ddot P > 0$). I investigate these intriguing results here.

There are several proposed models for QPE sources,  (Table 2 of King, 2023a lists 13 published models, for example repeated collisions of orbiting stars with a pre--existing accretion disc around the black hole), but the apparently universal preference 
for low--mass galaxy hosts with only moderate--mass ($M \sim 10^5 - 10^6\msun$) central black holes (hereafter `MBH')  strongly suggests that the possibility of partial tidal stellar disruption must play a role (cf King, 2023b). The observed quasiperiods and luminosities are consistent with a picture (King, 2020; 2022; 2023a,b)
where the star being disrupted is a white dwarf bound to an MBH in a very eccentric orbit, losing orbital angular momentum to gravitational radiation\footnote{It is likely that the same partial disruption process occurs for main--sequence stars, but these fill their tidal lobes at significantly larger separations, giving much lower accretion luminosities (see King, 2022, eqs 38, 39)}. 
This idea appears to be fairly successful both in describing the observed properties of all known QPE systems (including the predicted CNO enhancement in at least one case (Sheng, 2021)), and extending the sample to include other objects not previously recognised as QPE systems, such as HLX--1, with a period $\sim 1$~yr (see King, 2022). 

The very short quasiperiods ($18 - 7$~min) found by Masterson et al. (2025)
appear to make a WD donor unavoidable in 1ES~1927+654. Given its recent very rapid period evolution I note that if a more distant third star orbits the accreting BH--WD system, this can cause significant changes of the accretor's quasiperiod through von Zeipel--Lidov--Kozai (hereafter ZLK) cycles (cf King, 2023a: for a recent review of ZLK cycles see Perets, 2025). 

\section{ZKL Cycles in QPE Sources}

In ZLK cycles the inner binary (the QPE system in our case) continuously exchanges its orbital eccentricity $e$ and inclination $i$ because of the outer perturbing star. This has the largest component of the full system’s total angular momentum, so its orbital plane remains unchanged on timescales of interest, while the plane of the inner binary either oscillates between two fixed inclinations (libration) or revolves continuously (circulation). The exchanges of $e, i$ obey the constraint
\be
(1 - e^2)^{1/2}\cos i \simeq C
\label{ZLK}
\ee
where $C$ is a constant set by the initial conditions. This means that ZLK cycles have little effect on the angular momentum component $J$ of the inner mass--exchanging binary orthogonal to its instantaneous plane. This is precisely the angular momentum whose decay through gravitational radiation losses drives mass transfer. So a ZLK cycle 
cannot change the tidal lobe of the donor star in the inner binary.

\section{Evolution of the Mass Transfer Rate and Quasiperiod}

These constraints force the inner binary to evolve its quasiperiod as
\be
P \propto a^{3/2} \propto (1-e)^{-3/2}, 
\label{P}
\ee
and its GR--driven mass transfer rate as
\be
-\dot M_2 \propto P^{-8/3}(1-e)^{-5/2}\propto P^{-1}
\label{mdot}
\ee
King (2023a). This relation is completely independent of the nature (e.g. WD or MS) of the donor star, and follows solely from the fact that this star fills its tidal lobe 
%%%%%%%%%%%%%%%%%%%%%%%%%%%%%%%
and loses orbital angular momentum to gravitational radiation. 

In stellar--mass triple systems, mass transfer may result from other causes such as nuclear expansion or magnetic braking of the companion. In these cases there is no effect on the accretion luminosity, and the ZLK cycle only affects the system's inclination.   
%%%%%%%%%%%%%%%%%%%%%%%%%%%%%%

\section{Light Curves}
The characteristic timescale for ZLK changes is
\be 
t_{\rm ZLK} \simeq \frac{8}{15\pi}\biggl(1 + \frac{M_1}{M_3}\biggr)\biggl(\frac{P_{\rm out}^2}{P}\biggr)(1 - e_{\rm out}^2)^{3/2}
\label{tZLK}
\ee
(Antognini, 2015), where $M_1 , M_3$ are the black hole and outer perturbing star
masses,  and $P_{\rm out}, e_{\rm out}$ are the period and eccentricity of the outer binary. 

Logarithmically differentiating (\ref{ZLK}) gives
\be 
\frac{e\dot e}{1- e^2} \simeq -\tan i\frac{{\rm d}i}{{\rm d}t}
\ee
and so from (\ref{P})
\be
\frac{\dot P}{P} = \frac{3}{2}\frac{\dot e}{1 - e} \simeq -\frac{3(1 + e)}{2e}
\tan i\frac{\rm{d}i}{\rm {d}t}.
\ee
From equation (\ref{ZLK}) we find that at $i = 0$ (i.e. when the planes of the inner and outer binaries coincide) $e$ is maximal, and from (\ref{P}), $P$ is also maximal. Then from (\ref{mdot}) the mass transfer rate $-\dot M_2$  reaches a minimum, and from (\ref{tZLK}) the timescale $t_{\rm ZLK} \propto P^{-1}$ is also at a minimum.

This shows that 
the effect of ZLK cycles on a QPE system is to produce relatively brief minimum flux states in which the quasiperiod is long.  These minima are separated by long plateaux where the system is at maximum brightness and has a shorter quasiperiod. 
The system is maximally eccentric during its low states, and more circular during the plateaux. 
%%%%%%%%%%%%%%%%%%%%%%
Perets (2025) (Fig. 1) gives an explicit example of this type of cycle. Evidently the detailed shape of the light curve must depend on the precise values of the masses and initial conditions. 
%%%%%%%%%%%%%%%%%%%%%

\section{Application to 1E~1927+654}
King (2022) gives the general equations for eccentric QPE systems containing white dwarf donors. Using these and the observed period range from 18.1 to 7 minutes in 1E~1927+654 shows that the eccentricity 
varies between $e = 0.68$ in the bright plateaux, and $e = 0.97$ in the luminosity minima. The donor WD mass in the QPE `binary' is self--consistently constant at $M_2 \simeq 0.49\msun$ during the cycles. 
%%%%%%%%%%%%%%%%%%%%%%%%%%%%%%%%%%%

From (\ref{ZLK}) with $i = 0$ we find $C = (1 - 0.97^2)^{1/2} = 0.243$.
Then at the bright plateaux, where $e = 0.68$, we find from eq (\ref{ZLK}) that $ i = 74^{\circ}$,
which is the the angular excursion of the QPE binary above and below the orbital plane of the perturbing star.

%%%%%%%%%%%%%%%%%%%%%%%%%%%%%%%%%%%

This interpretation places interesting constraints on the nature of the perturbing star (of mass $M_3$). For a black hole mass $\sim 10^6\msun$ (Masterson et al, 2025) and values $M_3 \sim 1 - 10\msun$, a timescale (\ref{tZLK}) of order years requires either outer periods $P_{\rm out} \sim 10$~min, or an orbit which is also significantly eccentric. But periods $P_{\rm out} \sim 
10~{\rm min}\lesssim P$ are ruled out as they would produce a system with much more chaotic behaviour than what is observed. The eccentricity required for the perturber's orbit with a period significantly longer than the inner binary's 18 min 
to give $t_{\rm ZLK} \sim 1$~yr  is $e_{\rm out} \sim 0.98$, so rather similar to the maximum eccentricity of the QPE binary itself. This indeed suggests a similar dynamical origin. I note that the significantly longer orbital period means that this outer `binary' will outlive the QPE system. 
%%%%

%%%%%%%%%%%
%It appears likely that the currently observed triple system is a relatively stable remnant of a complex tidal capture event of significant duration involving several infalling stars. Evidence for this is an earlier changing--look event (Traktenbrot et al., 2019). There is no reason to assume that the dynamics of this event have finished

%and must evolve rapidly under gravitational radiation losses

%and dynamical interactions

%It seems likely that 1E~1927+654 will repay continued observation.

\section{Conclusion}

%The work of this Letter suggests that 1E~1927+654 is likely to continue to evolve rapidly. A test of the idea of ZLK cycles is that the mass transfer rate, and so probably the accretion luminosity, should vary as the inverse of the observed quasiperiod. 
%%%%%%%%%%%%
%%%%%%%%%%%
It appears likely that the triple system currently observed in 1E~1927+654
is a relatively stable remnant of a complex tidal capture event of significant duration involving several infalling stars. Evidence for this includes an earlier changing--look episode (Traktenbrot et al., 2019): the authors of that paper pointed out that 
the light curve was similar to those of tidal disruption events, but noted that the spectrum of the event did not support this conclusion. There is good reason to expect that the dynamics of the full event are far from finished, and the work of this Letter suggests that 1E~1927+654 is likely to continue to evolve rapidly and repay future monitoring. 

A test of the idea of ZLK cycles is that the mass transfer rate, and so probably the accretion luminosity, should vary as the inverse of the observed quasiperiod. 

It is perhaps worth emphasizing that, for this QPE system at least, there can be very little doubt that the donor is a white dwarf.
%%%%%%%%%%%%%%%%

%\section*{Acknowledgments}
% Lense--Thirring, capture of a BINARY, 
%\section*{Acknowledgments}
\section*{DATA AVAILABILITY}
No new data were generated or analysed in support of this research.
%I thank Phil Uttley, Adam Ingram, Rhanna Starling and Andrew Blain for stimulating discussions. I am very grateful to the referee for a perceptive 
%report.
\section*{ACKNOWLEDGMENTS}
It is a pleasure to thank Megan Masterson and Erin Kara for very helpful discussions. I wish also to honour the memory of my friend and colleague Jean--Pierre Lasota, with whom I discussed mass transfer processes in close binary systems over many years.

{}


\begin{thebibliography}{}
\bibitem{}
Antognini, J.M.O., 2015, MNRAS, 452, 3610
%\section*{REFERENCES}

%\bibitem{} 
%Arcodia, R., Merloni, A., Nandra, K., et al., 2021, Nature, 592, 704
%, doi: 10.1038/s41586-021-03394-6


%\bibitem{} 
%Chakraborty, J., Kara, E., Masterson, M., et al., 2021, ApJL, 921, L40 
%, doi: 10.3847/2041-8213/ac313

\bibitem{}
Chen, X., Qiu,Y., Li, S., Liu, F.K. 2022, ApJ, 930, 122

%\bibitem{}
%Collin--Souffrin, S., Dumont, A.M., 1990, A\&A, 229, 292
%\bibitem{}
%Cufari, M., Coughlin, E.R.,  Nixon, C.J., 2022,
%ApJ, 929, L20
%\bibitem{}
%Cufari, M., Coughlin, E.R., Nixon, C.J., 2022, ApJ, 929, L20

%\bibitem{}
%Farrell S. A., Webb N. A., Barret D., Godet O., Rodrigues J. M.,
%2009, Nature, 460, 73

%\bibitem{}
%Frank, J., King, A.R., Raine, D.J., 2002, {\it Accretion Power in Astrophysics}, 3rd ed.,
%Cambridge University Press

%\bibitem{}
%Gierli\`nski, M., Middleton, M., Ward, M., Done, C., 2008
%Nature, 455, 369
%\bibitem{} 
%Giustini, M., Miniutti, G., \& Saxton, R. D., 2020, A{\&}A,
%636, L2 
%, doi: 10.1051/0004-6361/202037610
%\bibitem{}
%Hills, J. G. 1988, Nature, 331, 687
%\bibitem{}
%Ingram, A., Motta, S.E., Aigrain, S., Karastergiou, A., 2021,
%MNRAS, 503, 1703 
%\bibitem{}
%King, A.R., 1988, QJRAS, 29, 1
\bibitem{} 
King, A.R., 2020, MNRAS, 493, L120
%, doi: 10.1093/mnrasl/slaa020
\bibitem{}
King, A.R., 2022, MNRAS, 515, 4344
\bibitem{}
King, A.R., 2023a, MNRAS, 520, L63K
\bibitem{}
King, A.R., 2023b, MNRAS, 526, L31
%\bibitem{}
%\bibitem{}
%King, A.R., Kolb, U., Burderi, L., 1996, ApJ, 464, L127
%\bibitem{}
%King, A.R., Pringle, J.E., Livio, M., 2007, MNRAS, 376, 1740
%\bibitem{}
%King, A.R., Ritter, H., 1998, MNRAS, 293, L42

%King, A.R., Lasota, J.P., 2014, MNRAS, 444, L30

%\bibitem{}
%King, A.R., Lasota, J.P., Middleton, M.J., 2022 New AR, to appear

%\bibitem{}
%Lasota, J. -P.,  Alexander, T.,  Dubus, G., Barret, D., Farrell, S. A., Gehrels, N.   Godet, O.,  Webb, N. A., 2011, ApJ, 735, 89L
%\bibitem{}
%Lasota, J.-P. 2001, New Astronomy Reviews, 45, 449

%\bibitem{}
%Lin, L.C.C., et al., 2020, MNRAS, 491, 5682

%\bibitem{}
%Lu, Wenbin \& Quataert, E., 2022, arXiv 221008023  
%\bibitem{}
%Lissauer
\bibitem{}
Masterson, M., Kara, E., Panagiotou, C., et al., 2025, Nature 638, 370
%\bibitem{}
%Middleton, M.,  Done, C., Ward, M., et al., 2009, MNRAS, 394, 250
%\bibitem{}
%Metzger, B. D., Stone, N. C., Gilbaum, S., 2022, ApJ, 926, 101
%\bibitem{} 
%Miniutti, G., Saxton, R. D., Giustini, M., et al., 2019,
%Nature, 573, 381 
%\bibitem{}
%Miniutti, G.; Giustini, M.; Arcodia, R. et al., 2023,
%A\&A 670, 93
%, doi: 10.1038/s41586-019-1556-x
%\bibitem{}
%Nixon, C.J., King, A.R., Price, D., Frank, J., 2012, ApJ, 757L, 24
%\bibitem{}
%Nauenberg, M., 1972, ApJ, 175, 417

%\bibitem{}
%Payne, A. V., Shappee, B. J., Hinkle, J. T., et al. 2021, ApJ, 910, 125

%\bibitem{}
%Raj, A., Nixon, C.J., 2021, ApJ 909, 82

%\bibitem{}
%Payne, A. V., Shappee, B. J., Hinkle, J. T., et al. 2022, ApJ, 926, 142

%\bibitem{}
%Peters, P.C., 1964, Phys Rev. 136, B1224

%\bibitem{}
%Peters, P.C. \& Mathews, J., 1963, Phys Rev 131, 435
\bibitem{} Perets, H.B., 2025, arXiv: 2504.02939
\bibitem{}
Sheng, Z.,  Wang, T.,  Ferland, G., et al.,  
%Shu, X.,  Yang, C.,  Jiang, N., Chen, Y., 
2021, ApJ 920L, 25

%\bibitem{}
%Sun, L., Shu, X., Wang, T., 2013, ApJ 768, 167

%\bibitem{}
%Song, J. R., Shu, X. W., Sun, L. M., et al., 2020, A\&A, 644, L9

%\bibitem{}
%A. Tiengo A., Esposito, P., M. Toscani,4, 5 G. Lodato,4 M. Arca Sedda,6 S. E. Motta,7 F. Contato,8 M. Marelli,2 R. Salvaterra,2 A. De Luca2, 3  arXiv:2202.08478
\bibitem{}Traktenbrot, B., Arcavi, I., MacLeod, C.L., et al., ApJ 883:94
%\bibitem{}
%Wevers, T., Pasham, D. R., Jalan, P., Rakshit, S., Arcodia, R., 2022,
%A\&A 659, L2
%\bibitem{}
%Terashima, Y., Kamizasa, N.,  Awaki, H., et al., 2012, ApJ 752, 154

%\bibitem{}
%Xian, J., Zhang, F., Dou, L., et al.,
%2021, ApJ 921L, 32

%\bibitem{}
%Zalamea, I., Menou, K., Beloborodov, A. M. 2010, MNRAS,
%409, L25


%\bibitem{}
%Webbe, R, Young, A.J., 2023, to appear in MNRAS (arXiv:2211.10176)

%Linial, I., \& Sari, R., 2017, MNRAS 469, 2441L 
%\\
%Miniutti, G., et al., 2019, Nature 573, 381
%\\
%Paczy\'nski, B.  \& Sienkiewicz, R., 1981, ApJLett 248, L27
%\\
%Peters, P.C., 1964, Phys Rev. 136, B1224
%\\
%Peters, P.C. \& Mathews, J., 1963, Phys Rev 131, 435
%\\
%Ritter, H., 1988, A\&A 202, 93
%\\
%Sepinsky, J. F., Willems, B., Kalogera, V., \& Rasio, F. A., 2007, ApJ
%667, 1170
%\\
%Soltan, A, 1982, MNRAS 200, 115
\bsp
\end{thebibliography}
\end{document}